\documentclass[twocolumn,twoside,slac_two]{revtex4}
\usepackage{graphicx}
\usepackage{fancyhdr}
\def\slashchar#1{\setbox0=\hbox{$#1$}           % set a box for #1
   \dimen0=\wd0                                 % and get its size
   \setbox1=\hbox{/} \dimen1=\wd1               % get size of /
   \ifdim\dimen0>\dimen1                        % #1 is bigger
      \rlap{\hbox to \dimen0{\hfil/\hfil}}      % so center / in box
      #1                                        % and print #1
   \else                                        % / is bigger
      \rlap{\hbox to \dimen1{\hfil$#1$\hfil}}   % so center #1
      /                                         % and print /
   \fi}                                         %
\def\etmiss{\slashchar{E}_T}			%
\def\ptmiss{\slashchar{p}_T}			%
\newcommand\lsb{\tilde{b}_1}
\newcommand\lst{\tilde{t}_1}

\newcommand\lntr{\tilde{\chi}_{1}^0}
\newcommand\neutr{\tilde{\nu}}

\begin{document}

\title{Search for Squarks and Gluinos with the D0 detector}


%\author{S.A. Uzunyan\footnote{speaker}\footnote{On behalf of the D0 collaboration}} 
% \author{Sergey A. Uzunyan} 
% \affiliation{Department of Physics, Northern Illinois University, DeKalb, IL 60115, USA
% \\On behalf of the D0 Collaboration}
\author{Sergey A. Uzunyan\footnote{On behalf of the D0 collaboration}} 
\affiliation{Department of Physics, Northern Illinois University, DeKalb, IL 60115, USA}
%%\\On behalf of the D0 Collaboration}
%%
%

\begin{abstract}
We report on D0 searches for scalar quarks ($\tilde{q}$) and gluinos ($\tilde{g}$), the superpartners 
of quark and gluons, in topologies involving jets and missing transverse energies. 
Data samples obtained with D0 detector from $p\bar{p}$ collisions at a center-of-mass 
energy of 1.96~TeV corresponding to an intergrated luminosities of 1--4~fb$^{-1}$ were analyzed.
No evidence for the production of such particles were observed and lower limits on squarks and
gluino masses were set. 

\end{abstract}

\maketitle

\thispagestyle{fancy}
\section{Introduction}
Supersymmetric (SUSY) extensions of the standard model (SM) 
assign a bosonic superpartner to every SM fermion and vice-versa. 
The theory that consider only the SM particles, their corresponding superpartners,
and two Higgs doublets is known as the minimal supersymmetric standard model (MSSM)~\cite{mssm}).
The MSSM Lagrangian depends on 105 parameters that makes it difficult to apply for searches 
involving complex production and decay of the SUSY particles.
Analyses described in section~\ref{sect_squark_and_gluinos} were done in the  
the minimal supergravity model with a reduced set of the parameters (mSUGRA~\cite{msugra}).
In mSUGRA the sparticles masses and couplings are defined by a set of only five
parameters: $m_0$ and $m_{1/2}$, respectively the universal scalar mass and 
the universal gaugino mass at $\Lambda_{GUT}$; $A_0$,
the universal trilinear coupling at $\Lambda_{GUT}$; tan~$\beta$ the ratio of the 
vacuum expectation value of the two Higgs fields; and $\mu$, the sign of Higgs-mixing mass
parameter. Searches for stop and sbottom quarks (section~\ref{sect_stop_sbottom})
do not depend on SUSY parameters other then the masses of the sparticles. In all presented analysis
R-parity is assumed conserved and thus the lightest supersymmetric particle (LSP) is stable
and all SUSY particles are produced in pairs. 
%%%%%
\section{The D0 experiment}
The D0 experiment was proposed in 1983 to study high mass states 
and large $p_T$ phenomena in proton-antiproton collisions at the Fermilab Collider Complex.
During 1996-2001 both the Tevatron and the D0 detector \cite{d0_upgrade} were significantly upgraded.
In Run~II (which started March 2001) the Fermilab collider operates at an increased 
center-of-mass energy of 1.96~TeV and at higher instantaneous luminosity 
(in range from $1.6\times10^{32}$ cm$^{-2}$s$^{-1}$ 
in the beginning of Run II up to $2.8\times10^{32}$ cm$^{-2}$s$^{-1}$ after 2006). 
%at higher instantaneous luminosity raging from  1.6 \times  $2.8x10^{32} up to $2.8\times10^{32} cm{-2}s{-1} ater 2006 ) 
The D0 detector upgrade included new central tracking and new forward muon systems 
and an improved central muon system. In 2006 the D0 tracking system was upgraded and
the trigger system was improved. This report describes the analyses of the Run II D0 data sets 
collected between April 2002 and December 2008.
%%%%%%%%%%%%%%%%%%%%%%%%%%%%%%%%%%
\section{Searches for squarks and gluinos}
\label{sect_squark_and_gluinos}
These searches were done using the mSUGRA model ignoring the production of scalar top quarks.
Thus below the ``squark mass'' is the average of all other squarks. 
\subsection{Search for squarks and gluinos in the jets plus $\etmiss$ final state}
A search was made in 2.1 fb$^{-1}$ of data for the $\tilde{q}\tilde{\bar{q}}/\tilde{q}\tilde{q}/\tilde{q}\tilde{g}/ \tilde{g}\tilde{g}
 \rightarrow jets + \etmiss$.  The analysis explored $(m_0,m_{1/2})$ plane of mSUGRA parameters with
fixed values of $A_0 = 0$, tan~$\beta$=3  and  $\mu<0$.  

The minimum number of jets in the final state could be two 
($p\bar{p} \rightarrow \tilde{q}\bar{\tilde{q}} \rightarrow q\lntr\bar{q}\lntr$,
 $p\bar{p} \rightarrow \tilde{q}\tilde{q} \rightarrow q\lntr q\lntr$), if the 
squark mass is less then the mass of the gluino;
three ($p\bar{p} \rightarrow \tilde{q}\tilde{g} \rightarrow q\lntr q\bar{q}\lntr$), at $m_{\tilde{q}}\approx m_{\tilde{g}}$;
or four ($p\bar{p} \rightarrow \tilde{g}\tilde{g} \rightarrow q\bar{q}\lntr q\bar{q}\lntr$), if $m_{\tilde{q}} > m_{\tilde{g}}$.
%Analysis corresponding to each of these final states futher denoted as ``dijet``,''3-jet``, and ''gluino`` respecively.
For the dijet analysis events were required to pass an acoplanar dijet
trigger, and a multijet and $\etmiss$ trigger was used for selection of events with higher number of jets.
The jets transverse energies were required to be $E_T > 35$~GeV for the first three leading jets and $E_T>20$~GeV 
for the fourth jet. In addition, the $\etmiss > 40$~GeV, the leading jet acoplanarity $<165^{\circ}$, the azimutal angles
between the $\etmiss$ and first jet $\Delta\phi(\etmiss,jet_{1}) > 90^{\circ}$ and between the $\etmiss$ and second jet were
$\Delta\phi(\etmiss,jet_{1}) > 50^{\circ}$ were required in all analyses. A veto on isolated electrons or muons with
$p_T>10$~GeV was also applied in all analyses to reject backrounds events with a lepton from $W$ boson decay. 
In the dijet analysis a cut on the minimum azimutal angle $\Delta\phi_{min}(\etmiss,jets)>40^{o}$ between
the $\etmiss$ and any jet with $E_T > 15$~GeV was applied to suppress the QCD background.
The two final cuts on $\etmiss$ and $H_T=\sum_{jets}E_T$ were optimized for each of the analyses 
by minimizing the expected upper limit on the cross section in the absence of signal.

Table~\ref{tab:squarks_gluino} shows the number of data and background events after all
selections. The background is dominate by $W/Z$+jets and $t\bar{t}$ processes. 
No excess of data events were found against the SM predictions in all three analysis and the 95\% C.L. limits on squark-gluino production cross-section were set. The QCD contribution was found 
to be small and was not accounted in the limit calculations.
\begin{table}[ht]
\begin{center}
\caption{Number of data and background events after all selections for each of the analyses.}
\begin{tabular}{|c|c|c|c|}
\hline 
Analysis & ($H_T$,$\etmiss$)~GeV &Data & Background\\
\hline
 $>=$2 jets & (325,225) & 11 & 11.1$\pm$1.2$^{+2.9}_{-2.3}$ \\
 $>=$3 jets & (375,175) & 9  & 10.0$\pm$0.9$^{+3.1}_{-2.1}$ \\
 $>=$4 jets & (400,100) & 20 & 17.7$\pm$1.1$^{+5.5}_{-3.3}$ \\
\hline
\end{tabular}
\label{tab:squarks_gluino}
\end{center}
\end{table}

\begin{figure*}[t]
\centerline{
   \includegraphics[scale=0.40]{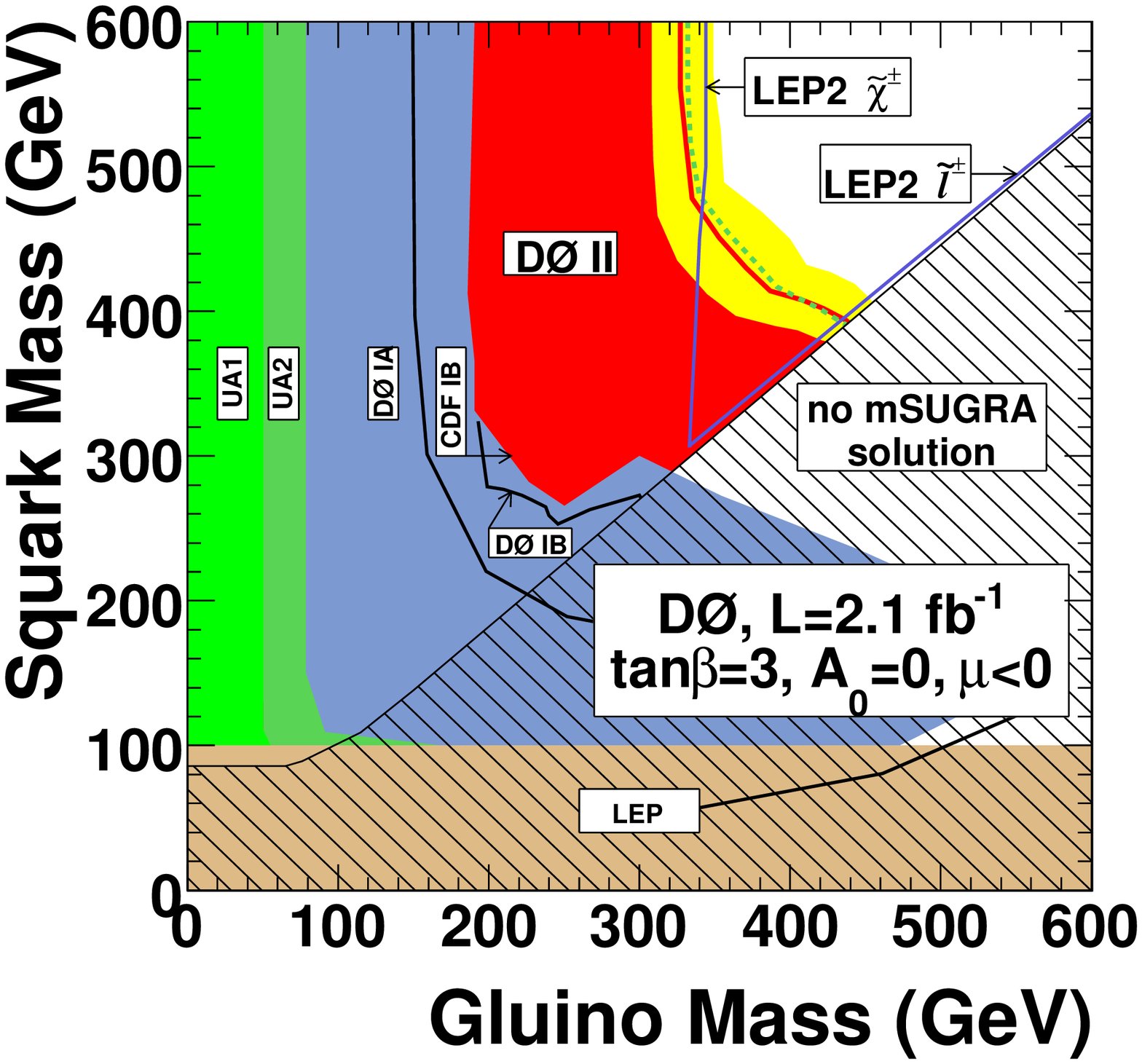}
   \includegraphics[height=75mm,width=95mm]{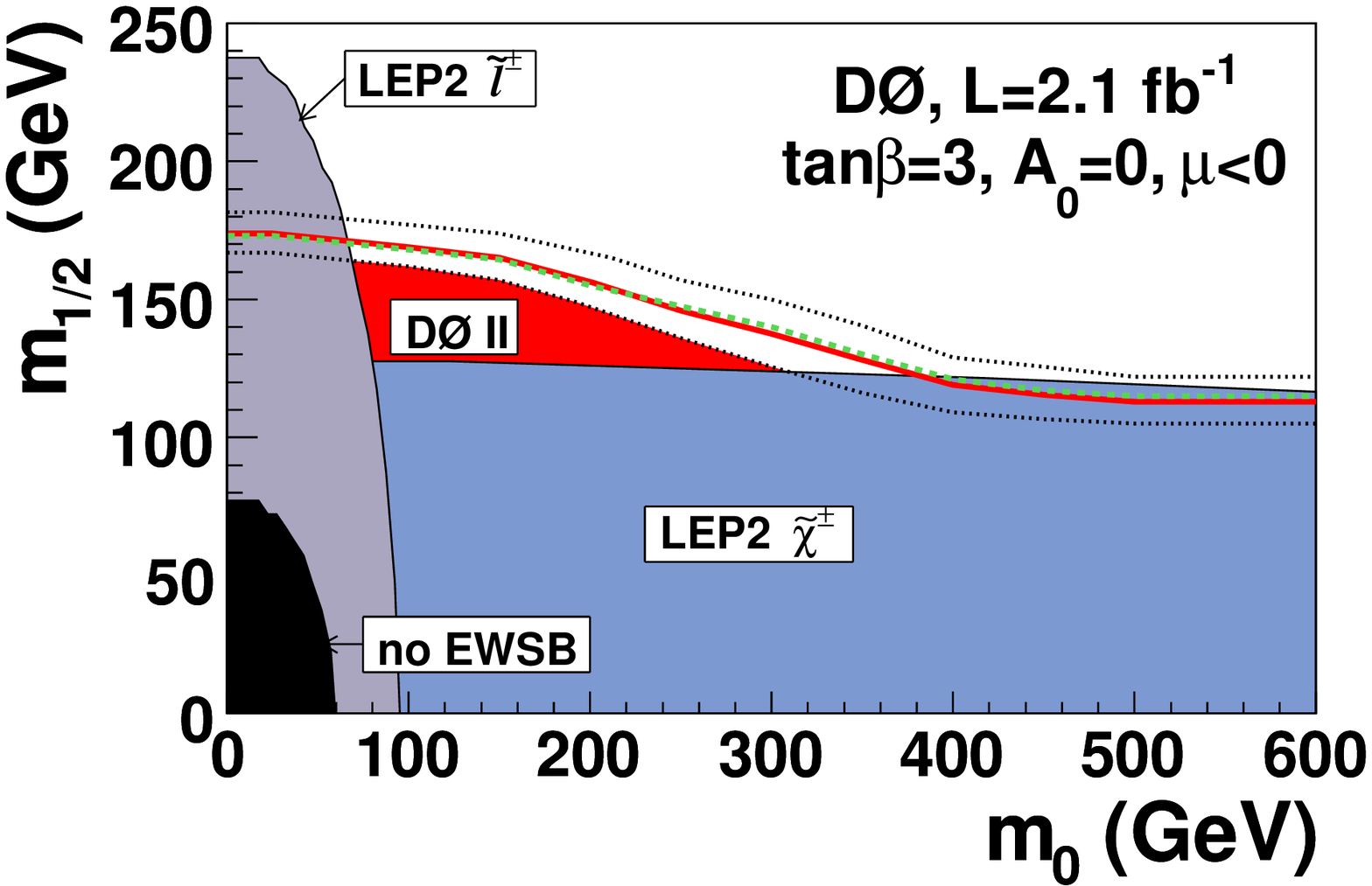}
}
 \leftline{ \hspace{2cm}{\bf (a)} \hfill\hspace{2cm} {\bf (b)} \hfill}
   \caption{\label{fig:exclusion_squarks_gluino_jets_met} The 95\%~C.L. excluded regions in the gluino and squark mass plane in the mSUGRA framework with tan~$\beta$=3, $A_0$=0, $\mu<$0 in the (a) gluino and squark mass plane and 
 (b) in the $(m_0,m_{1/2})$ plane. The thick (dotted) line is the limit of the observed (expected) excluded region 
 for signal cross section calculated with renormalization and factorization scale $\mu$= Q, with Q equal $m_{\tilde{q}}$,
 $m_{\tilde{g}}$, $(m_{\tilde{q}}+m_{\tilde{g}})/2$ for $\tilde{g}\tilde{g}$, $\tilde{q}\tilde{\bar{q}}$/$\tilde{q}\tilde{q}$ and  $\tilde{g}\tilde{g}$ production, respectively. The yellow band
 and the band delimited the two dotted lines correspond to the parton distribution functions and $\mu$ uncertainties.
 Red shading represent the D0 Run II results. Light shaded regions are excluded by previous experiments.}
\end{figure*}

Figure~\ref{fig:exclusion_squarks_gluino_jets_met}(a) show the excluded regions in the squark-gluino mass plane.
The observed  95\% C.L. limits on squark and gluino masses of 392~GeV and 327~GeV are the most constraining direct limits to date.
Figure~\ref{fig:exclusion_squarks_gluino_jets_met}(b) shows the excluded regions in the $(m_{0}, m_{1/2})$ mSUGRA parameter plane. The limits from LEP2 SUSY searches are improved for $m_0$ in the range 70--300~GeV and for $m_{1/2}$ values between 125 and 65~GeV. 
\subsection{Search for squarks production in the jets+tau leptons+$\etmiss$ final state}
A search explored the region of mSUGRA parameter space (tan~$\beta$=15, $A_{0}=-2m_{0}$,
$\mu<0$, $m_{0}$ and $m_{1/2}$ are varied) where squark and gluino cascade
decays could lead to final state with at least one $\tau$ leptons, two or more jets and
large missing transverse energy from the undetected neutralino. 
The complex cascade decays giving such a signature are only possible 
within a narrow region of ($m_{0}$,$m_{1/2}$) parameter space
and have not previously been studied at SUSY searches at Tevatron. 
The data sample corresponds to an integrated luminosity of 1~fb$^{-1}$.

The event selection are similar to the inclusive search and were optimized 
for ``tau-dijet'' or ``tau-multijet''
signatures depending on whether a dijet or multijet trigger were required. 
The analysis tuned to hadronically decaying tau leptons, 
which appear in the detector as narrow isolated jets.
Jets with transverse energy greater then 15~GeV were considered as tau-candidates 
if they were not already identified as one of the two highest $p_T$ jets. 
A neural net trained on simulated $Z \rightarrow \tau\bar{\tau}$ signal 
was used to separate tau candidates from quark and gluon jets. 
The signal selection was optimized (against the expected upper limit on the cross section)
with cuts on $\etmiss$ and $S_T = p^{jet_1}_T+p^{jet_2}_T+ E^{tau}_T$.  
After all selections 3 events were observed in data 
in good agreement with the SM prediction of $2.3\pm0.4(stat)\pm0.3(syst)$.
The background is dominated by top quark production and $W(\rightarrow l\nu)$+jets events. 
The QCD multijet backround contribution was found to be negligible.
Because the hadronic tau lepton can be reconstructed as a jet the obtained 
result was combined with the 2.1~fb$^{-1}$ dijet and multijet searches 
for squark production in the $\etmiss$+jets topology. 

\begin{figure*}[ht]
\centerline{
   \includegraphics[height=86mm,width=85mm]{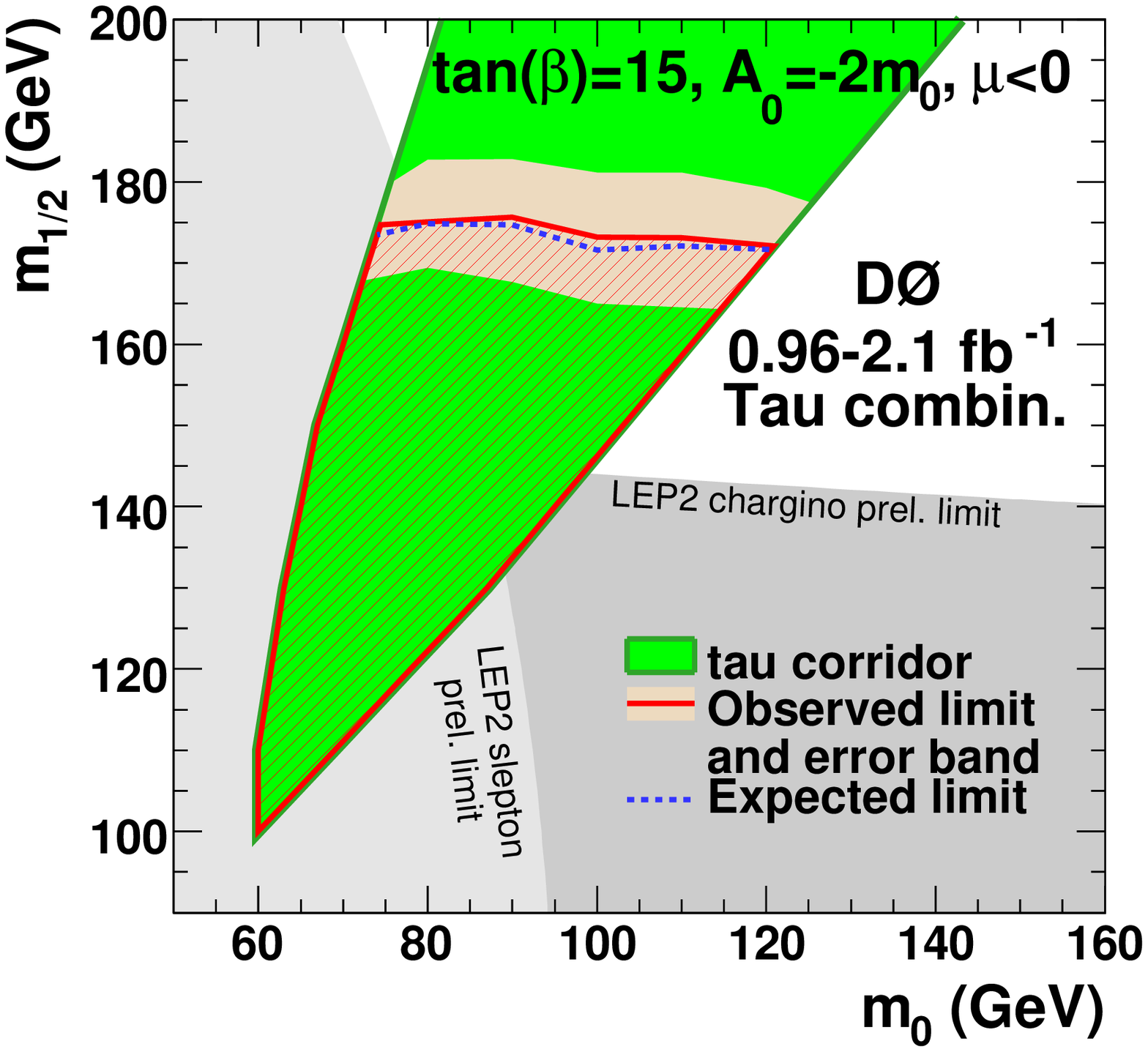}
   \includegraphics[height=76mm,width=90mm]{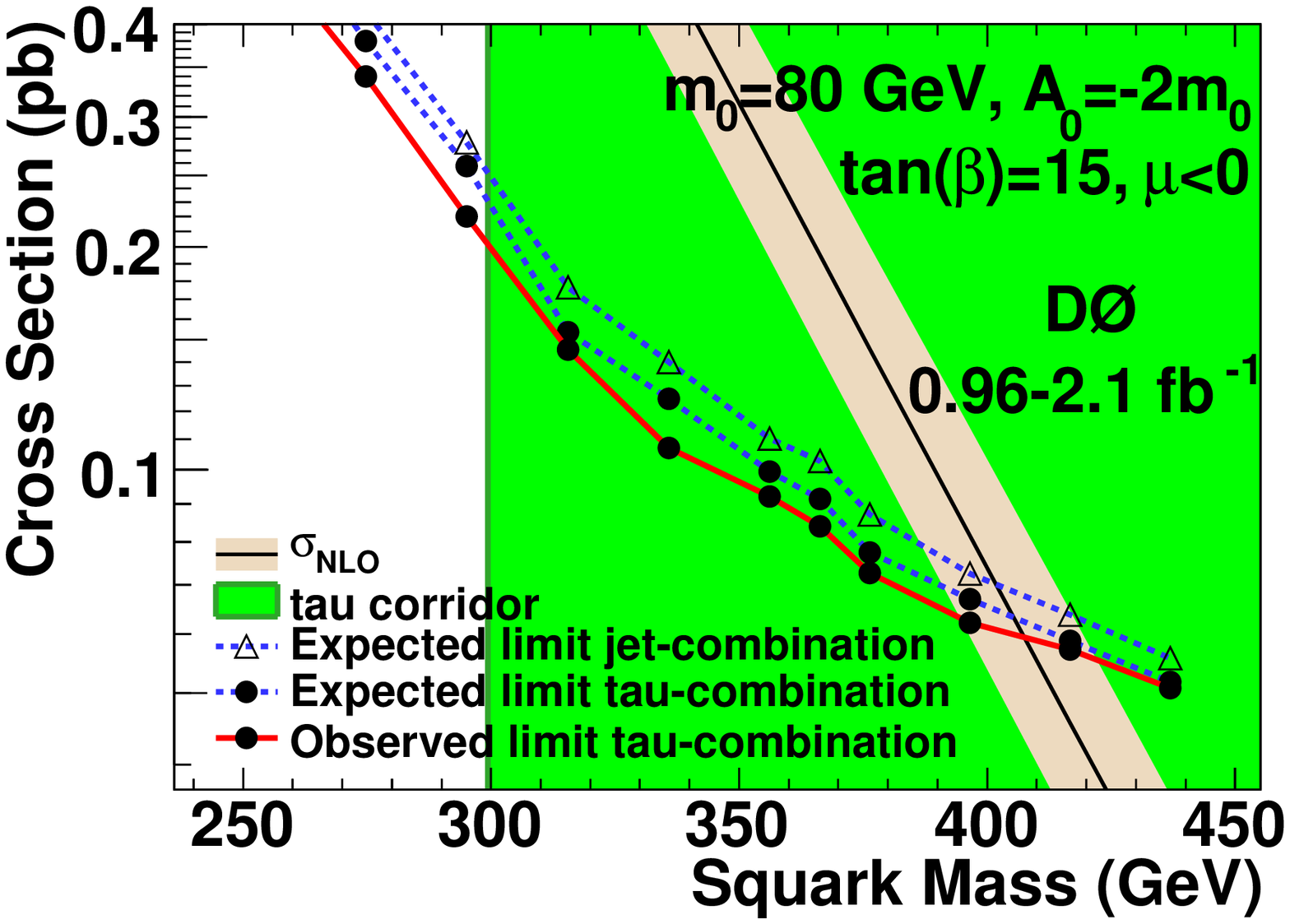}
}
 \leftline{ \hspace{2cm}{\bf (a)} \hfill\hspace{2cm} {\bf (b)} \hfill}
\caption{\label{fig:exclusion_squarks_gluino_jets_met_tau}a) The 95\%~C.L. excluded regions in the $(m_0,m_{1/2})$ plane
in the mSUGRA framework with tan~$\beta$=15, $A_{0}=-2m_{0}$, $\mu<0$.  The green region represents the  $(m_0,m_{1/2})$ space in which the final states with tau's are possible with the shaded part excluded by the combination of all squark analysis. 
 b) The 95\%~C.L. upper limits on squark and gluino pair production cross section for $m_0=80$~GeV. }
\end{figure*}

Figure~\ref{fig:exclusion_squarks_gluino_jets_met_tau}(a)  shows the expected and observed  
excluded regions in the $(m_0,m_{1/2})$ plane for the combination of all squark analysis. 
The result extends the exclusion region from LEP SUSY searches. The highest excluded squark mass at 95\% C.L. limits 
is 410~GeV, as illustrated in Figure~\ref{fig:exclusion_squarks_gluino_jets_met_tau}(b). 
%
%%%%%%%%%%%%%%%%%%%%%%%%%%%%%%%%%%
\section{Searches for pair production of the top and bottom squarks}
\label{sect_stop_sbottom}
The supersymmetric quarks (squarks) are mixtures of the 
superpartners $\tilde{q}_L$ and $\tilde{q}_R$ of the 
SM quark helicity states $q_L$ and $q_R$. The theory permits a mass difference
between squark mass eigenstates $\tilde{q}_1$ and  $\tilde{q}_2$ that 
gives the possibility that the lightest states of the stop and sbottom quarks are 
lighter than the first two generation squarks and thus have the largest
production cross section. The presented analysis also assumes
that the sneutrino (stop search) or the neutralino (sbottom analysis) is the LSP. 
%%%%%%%%%%%%%%%%%%%%%%%%%%%%%%%%%%
\subsection{Search stop pairs in 2 $b$-jets+$e\mu$+$\etmiss$ events}
\label{sect_stop}
A search was performed for three body decays of top squark pairs 
when each stop decays leptonically, $\tilde{t} \rightarrow b+\tilde{\nu}+l^{\pm}$ 
with a 100\% branching fraction. Thus the stop pair production 
final state corresponds to the $e^{\pm}+\mu^{\mp}+2b+\etmiss$ topology. 
The analysis studied 3.1~fb$^{-1}$ of data. 

The events were required to have an isolated electron and a muon of the opposite charge 
with $p_T$ greater then 15~GeV and 8~GeV respectively, 
accompanied by jets with transverse energy $E_T>15$~GeV and $\etmiss$.  
%Which triggers?
No explicit trigger requirements were applied, 
as the trigger efficiency is high for this combination of lepton $p_T$. 
The largest backrounds are $Z \rightarrow \tau\bar{\tau}$, QCD, WW, and top quark pairs. 
Cuts on angular differences between leptons and $\etmiss$ and $\etmiss>18$~GeV 
were applied to remove $Z$ and QCD events. 

The signal final state kinematic is determined by the size of the mass difference between squark and sneutrino    $\Delta(m_{\lst},m_{\neutr})$. Events with a large $\Delta(m_{\lst},m_{\neutr})$  have more
missing energy, higher jets energies and high charge leptons $p_T$, while signals with $\Delta(m_{\lst},m_{\neutr})$
have softer kinematics. The ``hard'' and  ``soft'' signal efficiencies after the $\etmiss$ cut are 15\% and 3\%
respectively. To prevent futher losses of the signal acceptances, a technique that
isolates the remaining background was applied: signal and backround  events 
were distributed into a two dimensional histogram in the $H_T$ (the scalar sum of the $p_T$ of the jets) 
and the $S_T$ (the scalar sum of $\etmiss$,  the electron $p_T$ and the muon $p_T$) plane.
The bin edges were selected such that the background events are concentrated into a few
bins while the remaining bins have a significant signal to background ratio.
%The bin edges were selected to concentrate the background events 
%into a few bins so that the remaining bins could have a significant signal to background ratio.
This method allows use of the same set of selections for the signals with different $\Delta(m_{\lst},m_{\neutr})$.

\begin{figure}[ht]
\centering
\includegraphics[width=80mm]{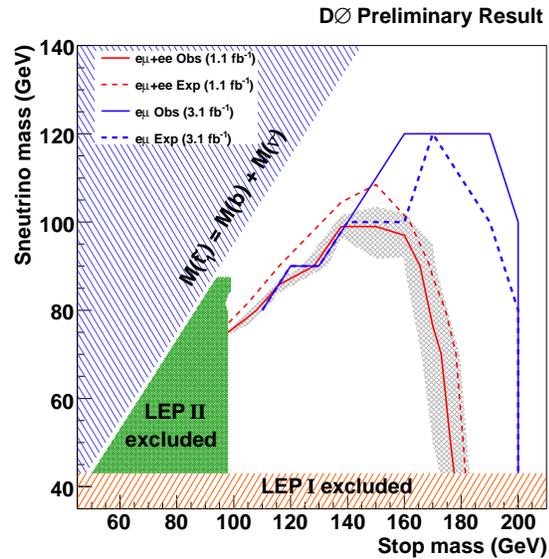}
\caption{Excluded 95\% C.L regions for the stop pair production in the dilepton plus $\etmiss$ events. 
Points below the blue lines are excluded by the presented 3.1~fb$^{-1}$ analysis. Also shown the D0 
1.1~fb$^{-1}$ exclusion countour from the $e\mu$ and $ee$ channels (the red line) and LEP I and LEP II excluded
regions (shaded green and orange regions). Dashed lines represent the expected exclusion contours.
  } 
\label{figure:stop_2jets_plus_emu_plus_MET_results_one_column}
\end{figure}

Table~\ref{tab:stop_pairs} shows the number of data, backgrounds and signal events after all selections.
No significant excess above the standard model predictions were found. Figure~\ref{figure:stop_2jets_plus_emu_plus_MET_results_one_column} shows the 95\% C.L. exclusion limits 
in the $(m_{\lst},m_{\neutr})$ mass plane. Stop masses up to 200~GeV were excluded for signals with $\Delta(m_{\lst},m_{\neutr}) > 30$~GeV and  $m_{\neutr} < 110$~GeV.
\begin{table}[h]
\begin{center}
\caption{Number of data, background and signal events after all selections for different stop signals.}
\begin{tabular}{|c|c|c|c|}
\hline ($m_{\lst}$,$m_{\neutr}$)~GeV     & Data & Background & Signal(acpt,\%)  \  \\
\hline
   (110,80)  &  288 & 303$^{+16}_{-20}$ & 89$^{+11}_{-13}$   (3) \\
   (150,50)  &  288 & 303$^{+16}_{-20}$ & 122$^{+14}_{-16}$  (15) \\
\hline 
\end{tabular}
\label{tab:stop_pairs}
\end{center}
\end{table}
%%%%%%%%%%%%%%%%%%%%%%%%%%%%%%%%%%
\subsection{Search sbottom pairs in 2 $b$-jets+$\etmiss$ events}
\label{sect_sbottom}
A search considered the region of SUSY parameter space where
the only possible decay of the sbottom quark's lightest state is
$\tilde{b}_1\rightarrow b\lntr$, $m_{\tilde{b}_1}>m_b+m_{\lntr}$ and
$m_{\tilde{b}_1}<m_t+m_{\tilde{\chi}_1^\pm}$,
where the neutralino $\lntr$ and chargino $\tilde{\chi}_1^\pm$ are the lightest 
SUSY partners of the electroweak bosons. These requirements result in $b\lntr\bar{b}\lntr$
final state. The corresponding detector signature are two acoplanar $b$-jets 
from the scalar bottom quarks and the missing energy due to escaping neutralinos. 
The data sample was collected using jet plus missing energy triggers 
and corresponds to an integrated luminosity of 4~fb$^{-1}$. 

Events with two or three jets with $E^{jet}_T > 20$~GeV and the $\etmiss>40$~GeV were selected.
For the two leading jets acoplanarity $< 165^\circ$ was required.
To reduce the contribution from $W\rightarrow l\nu$ decays, events 
with isolated electrons or isolated muons with $p_T > 15$~GeV were vetoed. 
To suppress the instrumental background events where 
the $\etmiss$ direction overlapped a jet in $\phi$ were removed.
Also events where the direction of $\etmiss$ is not aligned with the missing
track $\ptmiss$, calculated as the negative of the vectorial sum of 
charged particles transverse momenta were removed.
A neural net $b$-tagging tool was used to identify the heavy-flavor jets 
while significantly reducing the SM and QCD backgrounds which are dominated by light flavor jets. 
As the final state consist of two high $E_T$ $b$-jets,
the $X_{jj} \equiv  ( E_{T}^{jet1} +  E_{T}^{jet2} ) / (\Sigma_{jets} E_T)$ 
variable was used as a discriminant against top quark processes, requiring $X_{jj} > 0.9$.
Finally the cuts on $E_{T}^{jet1}$, $\etmiss$ and $H_{T}$ were optimized
for the different ($m_{\lsb},m_{\lntr}$) signals. These selections are tighter for the 
sbottom signals with large $\etmiss$. For the regions with a low $m_{\lsb}-m_{\lntr}$, the
 $\etmiss$ and jet energies in the signal events are lower and relaxed thresholds were used.

After all selections the number of data events agreed with the expectations 
from the simulated SM processes (dominated by the $t\bar{t}$ production and $W/Z$+heavy flavor jets events)
and the QCD estimation from data. The later is significant for the selections with soft requirement
on $\etmiss$.  Table~\ref{table_sbottom} shows the number of data, backround and signal events after all
selections applied in searches for the signals with low and high $\etmiss$. 
\begin{table}[h]
\begin{center}
\caption{Number of data,background and signal events after all selections for low and high $\etmiss$ sbottom signals.}
\begin{tabular}{|c|c|c|c|}
\hline ($m_{\lsb}$,$m_{\lntr}$)~GeV     & Data & Background & Signal(acpt,\%)  \  \\
\hline
   (100,60)  &  483 & 493$\pm$12 & 610$\pm$29    (1.0)    \\
   (240,0)   &  7   & 7.1$\pm$0.4 & 11.4$\pm$0.2 (3.6) \\
\hline
\end{tabular}
\label{table_sbottom}
\end{center}
\end{table}
\clearpage
\newpage

\begin{figure}[ht]
\centering
\includegraphics[width=80mm]{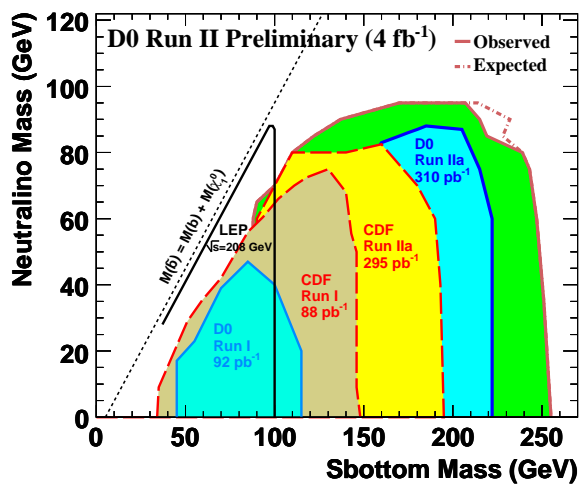}
\caption{The 95\%~C.L. exclusion contour in ($m_{\lsb}, m_{\lntr}$) mass plane. 
Also presented results from previous searches at LEP and at the Tevatron.} 
\label{figure:sbottom_2jets_plus_MET}
\end{figure}

Figure~\ref{figure:sbottom_2jets_plus_MET} shows the 95\%~C.L. and excluded region in the
sbottom mass versus neutralino mass plane. Production of sbottom quarks with $m_{\lsb} < 253$~GeV is 
excluded for $m_{\lntr} = 0$. The limits from previous Tevatron searches are improved and the exclusion 
region extends up $m_{\lntr} = 95$~GeV for $m_{\lsb}$ in the range 150--200~GeV.
%%%%%%%%%%%%

\section{Summary}
Searches for squarks and gluinos are performed in 1--4~fb$^{-1}$ D0 data samples for different 
regions of MSSM/mSUGRA parameter space. All presented analysis are in good agreement with
the SM predictions. As no signs of SUSY were observed a set of 95\% C.L. limits on squark
and gluino masses and on mSUGRA parameters have been obtained improving previous Tevatron results.
More details on the presented analyses can be found at~\cite{d0_np}.
%%%%%%%%%%%%%%%%%%%%%%%%%%%%%%%%%%
\bigskip % extra skip inserted

\end{document}